\documentclass[hidelinks]{scrartcl}
\usepackage{amsmath}
\usepackage[backend=bibtex,natbib=true,url=false,style=authoryear-icomp]{biblatex}
\usepackage{mathrsfs}
\usepackage{graphicx}
\usepackage[font=scriptsize,labelfont=bf]{caption}
\usepackage{hyperref}
\hypersetup{
	colorlinks = true,
    linkcolor = blue,
    urlcolor  = blue,
    citecolor = blue,
    anchorcolor = blue]
}
\usepackage{paralist}
\usepackage[textsize=footnotesize]{todonotes}

\bibliography{./biblio/weaver}

\begin{document}
\title{Spooky Action at No Distance}
\subtitle{\LARGE On the individuation of Quantum mechanical systems}
\author{\large David Weinbaum (Weaver) (\texttt{space9weaver@gmail.com})\\
\large The Global Brain Institute, VUB}
\date{\large \today}

\maketitle
\begin{abstract}
\footnotesize{Recent experiments have perfectly verified the fact that quantum correlations between two entangled particles are stronger than any classical, local pre-quantum worldview allows. This is famously called the EPR paradox first conceived as a thought experiment and decades later realized in the lab. We discuss in depth the nature of the paradox and show that the problematics it presents is first and foremost epistemological. After briefly exploring resolutions to the paradox that after many decades of discourse still remain controversial, we argue that the paradox is rooted in the failure of our current metaphysical scheme, being the foundation of our knowledge, to accommodate and cohere our knowledge of the phenomena of entanglement. We then develop and make the case for a novel and more fundamental resolution of the paradox by changing the underlying metaphysical foundation from one based on individuals to a one based on individuation. We discuss in detail how in the light of this new  scheme concepts central to the paradox such as realism, causality and locality are adjusted to the effect that the paradox is resolved without giving up these concepts so fundamental to our thinking. We conclude with a brief note about the important role of metaphysics to the progress of knowledge and our understanding of reality. \\
	
\raggedleft{\textbf{Keywords:} quantum entanglement, EPR paradox, individuation, metaphysics, realism , locality, causation}} 	

\end{abstract}
\pagebreak
\section{Introduction}
Every year the prestigious web magazine Edge\footnote{See: \href{http://edge.org/response-detail/26790}{http://edge.org/response-detail/26790}.} pronounces a yearly question and invites distinguished thinkers from diverse disciplines to answer. The 2016 Edge question was: ``What do you consider the most interesting recent scientific news? What makes it important?''. My motivation for writing this paper came from reading quantum physicist professor Anton Zeilinger's\footnote{See: \href{http://edge.org/memberbio/anton_zeilinger}{http://edge.org/memberbio/anton\_zeilinger}.} answer to this question. A quote from Zeilinger's answer is in place:
\begin{quotation}
The notion of quantum entanglement, famously called “spooky action at a distance” by Einstein emerges more and more as having deep implications for our understanding of the World. Recent experiments have perfectly verified the fact that quantum correlations between two entangled particles are stronger than any classical, local pre-quantum worldview allows. So, since quantum physics predicts these measurement results for at least eighty years, what’s the deal?

The point is that the predictions of quantum mechanics are independent of the relative arrangement in space and time of the individual measurements. Fully independent of their distance, independent of which is earlier or later etc. One has perfect correlations between all of an entangled system even as these correlations cannot be explained by properties carried by the system before measurement. So quantum mechanics transgresses space and time in a very deep sense. \textit{We would be well advised to reconsider the foundations of space and time in a conceptual way.} (my emphasis)	
\end{quotation}

My goal in this paper is exactly this: a reconsideration of the conceptual foundations of realism and specifically of space in the light of the phenomenon of quantum entanglement. I will apply Simondon's theory of individuation and Bergson's conceptualization of space in order to reexamine what Einstein called ``spooky action at a distance'' and more specifically the notion of physical locality and its underlying metaphysical assumptions. The examination will lead first to developing and arguing for an alternative metaphysical scheme that coherently accommodates both quantum and classical phenomena. This alternative scheme adjusts our understanding of realism and will further make a case for developing a new non-conventional (and somewhat surprising) intuition about the reality of space and how it is represented. With the renewed conceptualizations developed, it is shown that nothing spooky is taking place in quantum entanglement; at least not spooky in the sense that Einstein meant. The resolution of the paradoxical action at distance responsible for innumerable sleepless nights of physicists and philosophers alike, is thus shown to be possible by adopting the alternative metaphysical scheme we develop here. With it we can finally cohere our knowledge about quantum entanglement with the rest of our knowledge about natural phenomena.

The second section gives a short description of the EPR paradox and the violation of Bell's inequalities. It will then discuss more in depth the meaning of this violation and the problem of non-locality and causation in quantum entangled systems. The section concludes by proposing a conceptual revision of the problem and sketches the metaphysical adjustments that are needed in order to resolve the paradox. The third section starts with a critique of Boole's conditions of possible experience. Next, it presents in brief Simondon's theory of individuation and the application of the concept of individuation to entangled systems. It then develops in depth the new conceptions of realism and causal explanation in the light of a metaphysical scheme based on individuation. The fourth section is dedicated to the individuation of space and how it reflects on entangled systems. It starts with general considerations regarding the concept of locality as it is currently understood and its limitations. It then explores Bergson's metaphysics of space and applies it to argue that in the case of entangled systems also space and locality are subject to individuation. By that the development of the alternative metaphysical scheme started in section \ref{sec:individuation} is completed. A discussion of the philosophical implications on understanding locality in entangled systems follows. The fifth and last section is a summary of the whole conceptual revision developed in the paper and how it resolves the paradox. It concludes with a short note on the role of metaphysical investigation in cohering our knowledge about reality.   

\section{The EPR paradox and the nature of quantum phenomena} \label{sec:EPR}
\subsection{A short account of the EPR experiment and Bell's inequalities} \label{subsec:EPRparadox}
Around 1935 a paper authored by Albert Einstein, Boris Podolski and Nathan Rosen  \citep{einstein_can_1935}, presented a thought experiment that claimed to demonstrate that the quantum wave function does not provide a complete description of physical reality. This has come to be known as the EPR paradox. The thought experiment was set to show that a measurement of the location and momentum of two entangled physical particles can be performed in a manner that violates Heisenberg's uncertainty principle. Two physical particles are prepared in advance in such a manner that they are quantum entangled. The peculiar nature of quantum properties is such that for an entangled system of particles a certain property cannot be described or measured independently for each particle but only for the joint system as a whole\footnote{We will later see that this very peculiarity is central to the question of whether or not in the case of quantum entangled systems one can speak about two independent particles.}. Skipping the technical and mathematical details, the gist of the experiment was that once the entangled particles are physically separated in space, one can measure accurately the location of one particle and the momentum of the second. Since they are entangled, the momentum of the first particle can be accurately derived from the measurement of the second. This way one can measure both the location and momentum of one of the particles more accurately than what is allowed by the uncertainty principle. Two possible explanations are suggested: 
\begin{inparaenum}[\itshape a\upshape )]
\item Either the measurement performed on one particle affects instantaneously the other over an arbitrary distance to prevent the violation -- what came to be coined as ``spooky action at a distance''\footnote{Such effect however does not violate spacial relativity because no information is exchanged between the particles.}, or,  
\item the information about the outcome of all possible measurements is already present \textit{in both particles} to begin with and is encoded with some `hidden variables' that were set once the particles were brought into entanglement and carried along independently by each.	
\end{inparaenum}
To the authors, the possibility of non-local effects arising in entangled systems was unacceptable; which led them to the conclusion that the description of the entangled system of particles as a single non-decomposable system must be incomplete. In other words, the particles are singled out by hidden variables local to each. The incompleteness of quantum theory is that it falls short of predicting accurately the states of entangled particles though these are determined by their hidden variables. A proper local hidden variables theory that would presumably do better is needed to replace quantum theory. Such a hidden variable theory will affirm what is called \textit{local realism} also for quantum phenomena. Where in brief, locality basically means that no instantaneous action at a distance is possible and realism claims that physical particles possess definite properties irrespective to whether or when actual measurements are performed to obtain these properties. Thus a local realist quantum theory will cohere our knowledge about physical phenomena both classic and quantum under the same fundamental principles of locality and realism. The EPR experiment can be seen therefore as an attempt to `tame' quantum phenomena into an already established dogma.

In a seminal paper published in 1964, John \citet{bell_einstein_1964} came with a theorem stating that any physical theory that assumes local realism must satisfy certain conditions called Bell's inequalities. Bell developed a somewhat different version of the experiment described in the EPR paper using particle spins rather than location and momentum as the measured quantum properties. He showed that the predictions of quantum theory regarding measurements performed on entangled systems violate his inequalities. The consequences of such violation, if verified by actual experiments, will exclude any possibility of a local realist hidden variable theory to reproduce the results predicted by quantum theory.

Actual experiments equivalent to the EPR experiment were conducted since 1976 with overwhelming evidence that measurements of quantum entangled systems do violate Bell's inequalities. In the course of research various loopholes were discovered in the experiments and new setups were progressively devised to avoid them. In October 2015 the first loophole-free experiment was reported \citep{hensen_loophole-free_2015}, directly testing Bell's theorem and demonstrating yet again the peculiar nature of quantum phenomena. Probably it is this outstanding result that inspired Zeilinger's answer.  

There is a rich literature covering the complex physics of quantum entanglement which is far from being covered by the brief treatment given here. The important point however is that it has become finally evident by experiment that the phenomenon of quantum entanglement presents behaviors which are not coherent with our common intuitions regarding space and time. As Zeilinger concludes: ``Thus, it appears that on the level of measurements of properties of members of an entangled ensemble, quantum physics is oblivious to space and time.'' Trying to settle this apparent discrepancy and understanding the meaning of this radical statement is the point of departure of this paper.  

\subsection{Bell's inequalities and Boole's conditions of possible experience} \label{subsec:bell_ineq}
To develop a deeper understanding of Bell's inequalities and their meaning it is important to note that they do not describe a quantum physical principle or even a physical principle. Bell's inequalities rather state the conditions that must hold regarding knowledge that can be obtained by statistical sampling of a population of objects for which local realism holds. I.e. a population of objects that possesses (and are defined by) measurement independent properties and interact only according to the principle of locality. Pitowsky shows in \citep{pitowsky_george_1994} and more extensively in \citep{pitowsky_quantum_1989} that the Bell inequalities are a special case of \textit{Boole's conditions of possible experience}. Given a certain body of data concerning a population of objects, let $P_{1},P_{2},\ldots,P_{n}$ be the probabilities given of certain events. And where an event can be understood as the existence or non-existence of a certain set of properties in a single object. $P_{i}$ therefore is the frequency of finding a set of properties $i$ in the population. In the trivial case, where no relations obtain among the events, then the only constraints imposed on the probabilities is that $0\le P_{i}\le1$. 

However, if the events are logically connected, there are further equalities or inequalities that obtain among the different probabilities. Let us consider a simple example: suppose we get an urn with many balls some of which are wooden (event $E1$) with probability $P_{1}$ and some of them are red (event $E2$) with probability $P_{2}$. Now, we sample balls from the urn and we are interested in the frequencies of events such as $E1$ or $E2$ but also in the frequency $P_{12}$ of sampling balls which are both wooden \textit{and} red (event $E1\cap E2$). These three events here are not logically independent of course and in addition to the trivial inequalities $0\le P_{1},P_{2},P_{12}\le 1$ we also have: $P_{1}\ge P_{12}$, $P_{2}\ge P_{12}$. Also the frequency $P_{1}+P_{2}-P_{12}$ of sampling balls which are either wooden \textit{or} red (event $E1\cup E2$) must hold: $P_{1}+P_{2}-P_{12}\le1$. The various versions of Bell's inequalities\footnote{The Bell inequalities involve three primitive properties and their combinations. See \citep[pp. 103-104]{pitowsky_george_1994}} and other similar sets of constraints that are used in quantum theory are obtained in a similar way by applying logical rules to probabilities of properties and events (the occurrence of a single property is the simplest kind of event). Boole \citep{boole_theory_1862,hailperin_booles_1986} called these constraints conditions of possible experience because any observation/experience that involves probabilistic sampling of real properties of objects must logically stand to these conditions.

Remarkably, none of Boole's conditions can be violated when all the relative frequencies are measured \textit{on a single sample of the population}. In the above example suppose that we take 100 balls out of the urn. we discover that 60 are wooden and 75 are red and 32 of them are both red and wooden. In terms of frequencies, $P_{1}=0.6$ and $P_{2}=0.75$ and $P_{12}=0.32$. But then $P_{1}+P_{2}-P_{12} > 1$ which is a logical impossibility because in such case there must be a ball which is `red', `wooden' but not `red and wooden'. Without exception, as long as we make all measurements on a single sample, similar logical impossibilities will arise in conjunction with the violation of one or more of Boole's conditions also in arbitrarily complicated cases. However, if for some reason or other, the measurements of logically dependent events are not made on a single sample, violations of Boole's conditions may occur. \Citet[pp. 105-107]{pitowsky_george_1994} lists a few reasons why such violations may happen:

\begin{description}
	\item[1.~Failure~of~randomness] -- A violation of Boole's conditions may occur if one or more of the  distinct samples fail to represent the distribution properties in the overall population. The population might be not well mixed, the samples too small etc.
	\item[2.~Measurement~biases] -- Even if the samples are perfectly random, violation can still occur if the observations are somehow biased or disturbed because they are not well performed. In the above example we could imagine that the property 'red' is observed under certain lightning conditions while the property 'red and wooden' is performed under different lighting conditions. 
	\item[3.~No distribution] -- According to the law of large numbers, the relative frequency of a property in a finite random sample approximates, with high probability, the frequency of that property in the larger population. For that reason we expect Boole's conditions to hold even when relative frequencies are measured over distinct samples. But this consideration hides the assumption that the hypothetical population we examine does have an \textit{a priori} distribution of properties that is \textit{the cause} for the measurements obtained. But according to Hume's empiricist skepticism \citep{hume_treatise_2012}, the attribution of causal explanation between two events cannot be logically justified. Specifically, the attribution of relative frequencies to an \textit{a priori} distribution of the population is merely an induction. The failure of Boole's conditions may therefore arise even when samples are sufficiently randomized and measurement biases have been eliminated simply because the habitual assumption about a causative explanation is not valid. It might well be that there is no population with stable properties and consequently no distribution. There is also the case that somehow properties do not exist independently of measurement. This consideration is not merely technical like the first two and will be further discussed later.
	\item[4.~Mathematical~oddities] -- Within certain mathematical considerations of how the probability measure is defined, there are exotic cases of distribution in a continuous probability space ( as opposed to discrete populations of objects) where there is a logical possibility of the violation of Boole's conditions. But this option will not be our concern here as it can hardly apply to natural phenomena.   
\end{description}
The nature of quantum phenomena is such that certain sets of properties are complementary (e.g. the position and momentum of a particle) which means that according to the uncertainty principle, these properties cannot be measured simultaneously, or, at least we do not know how to perform such simultaneous measurements. Still, these properties do hold between them logical dependencies such as those discussed here. Interestingly, Pitowsky notes that when we do know how to perform simultaneous measurements of a certain set of properties (i.e. in the case they are not complementary), there is no violation of Boole's conditions even when the measurements are performed on distinct samples \citep[p 109]{pitowsky_george_1994}. However, testing the behavior of quantum entangled systems specifically involve complementary properties for which no method for simultaneous measurement is known to exist. In such cases, it is necessary to perform measurements over multiple distinct samples. It is in such cases and only in such cases that quantum theory predicts outcomes that are in violation of Boole's conditions. A violation of Boole's conditions therefore is unique to those cases where no method for simultaneous measurement is known to exist.

This analysis clearly exposes the serious threat that quantum phenomena presents to the consistency of our knowledge and understanding of natural phenomena. The paradox goes beyond physics and one needs therefore to come up with a convincing explanation for the violation of Boole's conditions. In other words, to plausibly show that the arising logical contradictions involved are only apparent, that there is no actual threat to our logical conceptions. Of the four categories of explanations mentioned above, we can quite safely discard the fourth category of mathematical oddities as a relevant candidate if only because of Ockham's razor \citep[p 119]{pitowsky_george_1994}. In regard to the first category -- failure of randomness, this is basically a technical issue that early experiments were ridden with. New measurement methods and the progressive elimination of various loopholes as reported in \citep{hensen_loophole-free_2015} pretty much eliminates this category too. We are left with the second and third categories which are more interesting because in contrast they offer explanatory interpretations of quantum mechanics that challenge our most basic intuitions.  

\subsection{The problem of non-locality and realism in entangled systems} \label{subsec:realism}
The second category of explanations involve measurement biases of two kinds: 
\begin{inparaenum}[\itshape a\upshape )]
\item a bias that depends on the measurement equipment and method and that can be eliminated by improved technology, and
\item a bias which is built-in in the setup of experiments that cannot be removed by improved technology.  
\end{inparaenum}  
Explanations that involve measurement biases of the second kind are usually referred to as hidden variables theories. It is hypothesized that such variables that are not defined in the current quantum theory (i.e. hidden form it) display dynamic changes that bias the results of measurements in such manner as to produce  \textit{the appearance of violation} of Boole's conditions. In other words, the violation is an illusion only indicating the incompleteness of the current theory. Were we in possession of a better theory that exposes the hidden variables, its predictions wouldn`t have violated Boole's conditions at all. As already discussed in subsection \ref{subsec:EPRparadox}, experimental reality is quite embarrassing in this respect: all hidden variable explanations coherent with experiment involve non-locality i.e. effects that propagate instantaneously across arbitrary distances. Again, it is not that something is fundamentally wrong with hidden variables explanations; they simply do not add anything that explains away the disturbing peculiarities of quantum entanglement. Instead, they just re-describe them in different terms leaving our deepest intuitions about locality in question. Still, having to choose between logical contradiction and non-local effects, the latter is the lesser evil. 

We are left to consider the third category of explanations. As was already discussed above, the third category brings up the problem of realism. Given the violation of Boole's conditions in the case of measuring complementary quantum properties, and given the inconvenience invoked by non-locality as a possible justification, we may consider the alternative that an \textit{a priori} distribution of the population of events/objects under examination does not exist and if so, distinct samples \textit{do not represent} a single hypothetical population in possession of stable properties that exist independently of measurement or other interactions. As Pitowsky puts it:
\begin{quotation}
	What is at stake is the idea of causality. The 'no distribution' approach takes the view that certain phenomena,  or more precisely, certain aspects of certain phenomena, have no causal explanation. They simply occur and that is it. 
\end{quotation}
This approach is not a radical departure from the general empiricist suspicious view of causal explanations. The function of a scientific theory, as perceived by the empiricist, is to organize data and predict. Causal explanation is a fiction of the human mind riding on a theory's ability to organize and predict. Which intuition locality or realism would be worth keeping and which could be sacrificed to make the behavior of quantum entanglement a `possible experience'? At this point this is a matter of controversy among both physicists and philosophers.

The philosophical riddle presented here is apparently an epistemological one. A fundamental part of the physical world - the world of the very small, behaves in a way that seems to put in doubt the human ability to create a unified and coherent corpus of knowledge. We cannot do with 'impossible experiences' running havoc in our laboratories. Given the history of research in quantum physics it is not very plausible (though not ultimately refutable) that some fine detail of the theory has escaped us and once it will be discovered, everything will be put in order. The controversy regarding non-locality and realism that remains unresolved is philosophically very disturbing. In as far as the empiricist physicist is concerned, physics is okay; quantum theory is one of the most successful scientific theories ever devised. It is our conceptions that need revision.

\subsection{A conceptual revision}
The currently accepted idea is to either give up locality or realism. As each alone seem to resolve the paradox, it would seem reasonable that choosing only one minimizes the 'damage' inflicted on our sensibilities. But perhaps there is a way to somehow give up both in their current form and instead rethink their deeper meaning in a way that will shed new light on possible experiences and will allow us to keep both albeit with a slight yet profound new meaning. In the following I am going to describe and defend such an alternative approach. We start with the claim that the paradox we are facing is not merely epistemological but is rooted in the very concept of space insofar it is applied to locality and in what we conceive as real insofar it is applied to the properties of physical entities. In other words, the impasse the paradox presents is metaphysical. It is metaphysics that shapes our intuitions and it is metaphysics that needs to be adjusted. Here is a sketch of the proposed adjustment: 
\begin{enumerate}
\item  In developing Boole's conditions of possible experience, Aristotle's principle of the excluded middle is implicitly taken as given. That is, a property either exists or does not exist in any object or event of interest. Our criticism of Boole's conditions is that while this assumption may be legitimate for abstract objects and properties, its automatic extension to physical objects and events is far from warranted. Independently of one's knowledge about a certain property, it is conceivable that properties undergo a process of genesis or differentiation and are not \textit{a priori} given or just appear instantaneously.\label{item:1}
\item Simondon's theory of individuation proposes a metaphysics of formative processes that replaces the metaphysics based on fully formed individuals on which Boole's conditions are based.  The idea of individuation allows to replace the\textit{ hard realism} described by Boole's conditions with a \textit{soft realism} where properties and entities defined by properties are not given \textit{a priori} as fully formed individuals but undergo a process of coming into being -- individuation. The violation of Boole's conditions when applied to undifferentiated properties then merely indicates the inadequacy of hard realism as a description of quantum phenomena whereas soft realism is entirely consistent with it. \label{item:2}
\item In subsection \ref{subsec:individuation_realism}, soft realism is shown to be a position midway between the commonly accepted hard realism and the non-realist position discussed in subsection \ref{subsec:realism}. Replacing hard realism with soft realism and individuals with individuation carries profound consequences on understanding causation in quantum mechanical systems and brings us closer to a consistent understanding of entanglement (and other quantum effects as well).    \label{item:3}
\item The principle of locality considers spatial distinctions and effects over distances. If spatial distinctions are subject to processes of individuation like other physical properties, it is possible that our conception of the spatial separation at the basis of locality requires refinement. Such refinement is proposed by Bergson's metaphysics of space as will be discussed in subsection \ref{subsec:bergson}. \label{item:4}
\item  Based on step \ref{item:4}, it is argued that in the case of entangled systems, and prior to measurement, spatial separation and therefore distance as we conventionally conceive \textit{do not exist}. In other words, an entangled system while being spatially extended, still exists in a single and not yet divisible (individuated) locality. This is shown to be coherent with the adjusted understanding of causality discussed in subsection \ref{subsec:causality}. \label{item:5} 
\item  It will follow that the predictions of quantum theory can be made consistent with possible experiences without giving up neither realism nor locality on condition that we ground possible experiences on the metaphysics of individuation proposed here.
Since the metaphysics of individuation does not exclude individuals, it seems to work remarkably well in cohering our understanding of both classical and quantum phenomena.  \label{item:6} 		  	
\end{enumerate}

\section{Individuation and its application to physical systems} \label{sec:individuation}
\subsection{Critique of Boole's conditions of possible experience}  % Covers step 1
As already discussed in subsection \ref{subsec:bell_ineq} Boole's conditions of possible experience arise as a combination of logical propositions about properties of objects or events and the probability of observing combinations of such properties. In the discourse to this point, resolving the apparent paradox of the violation of Boole's conditions was a matter of providing physical or technical interpretations. Yet, there is another, less obvious, option: that Boole's conditions themselves are the problem. What if, contrary to our common-sense assumptions, Boole's conditions are not the proper method of universally representing possible experiences? 

The notion of possible experience is quite profound; it makes explicit that in the world of phenomena, not anything goes. In other words, that in the interactions between an observer and the world certain regularities and conditions hold that make experiences appear coherent and consistent. Boole's conditions are in fact a metaphysical scheme representing a fundamental belief about how the world is and how it can be represented. Specifically that would mean:
\begin{inparaenum}[\itshape a\upshape )]
	\item the world can be described as a collection of objects, events and relations among them;
	\item objects and events are individuals defined by concrete sets of properties;
	\item individuals can be represented by predicates that specify their properties; and finally,
	\item individuals, their relations and modifications can be represented and reasoned about in terms of logical propositions about their properties.  	
\end{inparaenum}
Aristotle's principle of the excluded middle that a property cannot both exist and not exist at the same time and there is no third option (the middle)\footnote{Interestingly, in eastern philosophy, there is a third option and even more than one. See for example: \href{https://aeon.co/essays/the-logic-of-buddhist-philosophy-goes-beyond-simple-truth}{https://aeon.co/essays/the-logic-of-buddhist-philosophy-goes-beyond-simple-truth}}, establishes the individual as a consistent and coherent concept. 

In the light of the obvious violations of Boole's conditions, we criticize the universal adequacy of this metaphysical scheme. Perhaps there are phenomena that cannot be given as individuals and therefore their representation as individuals cannot be expected to yield logically consistent description of experience? In the urn example discussed in subsection \ref{subsec:bell_ineq}, a ball cannot be wooden, red but not \textit{wooden and red}. But even such common-sense example is warranted to work only as long as we deal with abstract representations of properties. In the actual world however, it is not free of problems and hidden assumptions and cannot be warranted to work in all cases\footnote{E.g. when the observations are made on distinct samples, conditions of lighting may affect the observed color. Also things remain wooden only within a definite temperature range that might change from sample to sample, etc.}.  

Another remarkable example that supports our critique is the questionable individuality of a lump of sixteen cells (prior to blastulation) developing from a human fertilized egg but prior to any differentiation. Is this merely a lump of cells, a human fetus (a human person with person rights), or a tiny part of the mother's body? It is entirely unclear how to categorize the object as the same set of properties can satisfy multiple sets of propositions, each with very different and far reaching consequences. In such cases of under-determined objects we have two options: either to add an additional metaphysical presumption (e.g. the idea of spiritual conception at the moment of fertilization) that will provide the missing determination, or, we can delay our answer and wait till a natural developmental process will provide further determination. 

It can be objected that in this example, the properties are given as facts and therefore this is not a real problem but a question of the interpretation of facts, but a deeper examination that will not be carried out here, can show that like in many other examples a complete separation of subsets of properties that will distinctly determine (identify) either case is not possible. Another objection would be that at any case we can never know everything about an individual and therefore our representations are inherently partial to the actual object being represented. There are always properties which are hidden from us and these, once known, will resolve any question of determining the nature of any phenomena in a consistent and coherent manner. This is indeed the claim of all hidden variable explanations in our case. The objection tries to explain away the metaphysical problem on the basis of the incompleteness of our knowledge. Clearly, in the case of quantum entanglement this explanation fails for even if we hypothetically had all the necessary facts still the paradox persists.      

In phenomena such as quantum entanglement it is not anymore the case that one could argue that separability and inseparability are only a matter of interpretation of the facts. The inseparability of entangled pairs is the fact of the matter which casts a profound doubt whether the metaphysical scheme of individuals is indeed universally fit to describe natural phenomena. Apparently, quantum phenomena cannot always be represented in terms of individuals, no wonder that Boole's conditions of possible experience may be violated. A reasonable response to this critique is that our metaphysical scheme must be adjusted and extended to account for those cases where experiences are given but cannot be represented in terms of individuals. The next subsection introduces an alternative metaphysical scheme that transcends individuals. 
                   
\subsection{Simondon's theory of individuation} % Covers step 2
To grasp the concept of individuation, we first need to briefly review how the metaphysical scheme based on individuals with an \textit{a priori} given, unambiguously defined, stable identity accounts for change and the genesis (individuation) of individuals. Generally speaking, we need to identify a principle(s) and the specific initial conditions of its operation that together bring forth the individual. For example, planet earth is an individual object. To account for its genesis, astrophysicists developed a theory about the formation of planets and the necessary conditions for planets to form, e.g. the existence of a star such as the solar system. Inasmuch as this scheme makes sense, it suffers a major weakness: it only shows how individuals (planets) are formed by positing other individuals -- in this case these are the identified necessary conditions that are given \textit{a priori} and an individual guiding principle -- a theory of planets formation. Clearly, in the very way we commonly think and represent the world, individuals are the primary metaphysical elements and individuation is only secondary \citep{weinbaum_complexity_2015}. It follows therefore that we must always assume an already fully formed individual prior to any individuation. 

Gilbert Simondon was the first to criticize in depth the classical treatment of individuation and the majority of his writings \citep{simondon_individuation_2005} are dedicated to developing a new philosophy of individuation. In \citep{simondon_position_2009} he explains: 
\begin{quotation}	
	``Individuation has not been able to be adequately thought and described because previously only one form of equilibrium was known--stable equilibrium. Metastable equilibrium was not known; being was implicitly supposed to be in a state of  stable equilibrium. [...] Antiquity knew only instability and stability, movement and rest; they had no clear and objective idea of metastability.'' (see ahead)
\end{quotation}

Simondon offers a metaphysical scheme where the process of individuation is primary while individuals are secondary products. The individual is only a relatively stable phase in a dynamic \textit{metastable process} and is always in possession of not yet actualized and not yet known potentialities of further individuation. He writes:
\begin{quotation}
	``Individuation must therefore be thought of as a partial and relative resolution manifested in a system that contains latent potentials and harbors a certain incompatibility within itself, an incompatibility due at once to forces in tension as well as to the impossibility of interaction between terms of extremely disparate dimensions.'' \citep{simondon_position_2009}
\end{quotation}
The process of individuation is described as the progressive determination of that which is determinable in a system but is not yet determined. Individuation is about the formation of distinctions that did not exist previously -- it is about differentiation\footnote{In its wider sense individuation speaks about both the formation and dissolution of distinctions.}. An individual therefore is not anymore a rigid unity with ultimately given properties but rather a plastic and dynamic entity in a metastable state punctuated by events of transformation. Every such event reconfigures the system and the manner by which further transformations become possible. 

\subsubsection*{Metastability}
The concept of metastability is central to Simondon's theory. A metastable system is a system with a number of temporary stable states where each state may display different properties. Driven by the occurrence of external perturbations, a metastable system moves among states of local stability and hence the designation that implies that no single state is ultimately stable. Furthermore, in metastable systems properties may differentiate or merge, distinct states appear and disappear and the very boundaries delineating the system may change. Metastability implies a tension between stable and unstable aspects of the individual \citep{combes_gilbert_2013}.  

\subsubsection*{The preindividual}
In its process of individuation, an individual is preceded by a state of affairs which is yet undetermined -- the \textit{preindividual}. Deleuze, whose seminal work \textit{Difference and Repetition} draws on many of Simondon's insights, would later describe the preindividual as  ``determinable but not yet determined'' and individuation basically proceeds as the preindividual's ``progressive determination''\citep{deleuze_difference_1994, weinbaum_complexity_2015}. The preindividual must not be understood as a kind of ultimate disorder. It may contain partially individuated entities and principles that instruct its evolution to some extend but the combination of whom cannot fully determine the outcome. Even after an individual has reached a relatively stable state or formation, the preindividual is not necessarily exhausted and keeps on persisting in the individuated system as a source of inherent instability. It is the presence of the preindividual that allows subsequent individuation. 

The unity characteristic of fully individuated beings (i.e. identities) and warranted by the application of the principle of the excluded middle, cannot be applied anymore to the preindividual. The preindividual is that intrinsic aspect of the individual that goes beyond its unity and identity. It is important to emphasize here the metaphysical sense in which this is said: \textit{individuals are not only more than what they appear to be (in our representations), but also more than what they actually are}. Precisely here lays the paradigmatic shift in the metaphysical scheme from \textit{being} (individuals) to \textit{becoming} (individuation).    

\label{relations} Simondon also emphasizes that relations between individuals undergo individuation too: ``A relation does not spring up between two terms that are already separate individuals, rather, it is an aspect of the \textit{internal resonance of a system of individuation}. It forms a part of a wider system.'' \citep[p. 306]{simondon_genesis_1992}. Furthermore, individuation never brings to light an individual in a vacuum but rather an individual-milieu dyad. This dyad contains both a system of distinctions and a system of relations. The individual and its milieu reciprocally determine each other as they develop as a system wider than any individual.

\subsubsection*{Transduction} \label{subsubsect:Transduction}
\textit{Transduction} is a technical term Simondon is using to designate the abstract mechanism of individuation. The term captures some of the most innovative (and important to our case) characteristics of individuation. Understanding the term cannot make use of classical logic and procedural descriptions because they require the usage of concepts and relationships among concepts that only apply to the products of the operation of individuation \citep[p. 10]{simondon_position_2009}. Transduction comes to designate therefore a metaphysical scheme that is constructed from a generative point of view that precedes any \textit{a priori} given individuals. In Simondon's words:
\begin{quotation}
 One could, without a doubt, affirm that transduction cannot be presented as a model of  logical procedure having the value of  a proof. Indeed, we do not wish to say that transduction is a logical procedure in the current sense of  the term; it is a mental process, and even more than a process, \textit{it is a functioning of  the mind that discovers} [emphasis added]. This functioning consists of  following being in its genesis, in carrying out the genesis of thought at the same time as the genesis of the object. \citep[p. 11]{simondon_position_2009}.
\end{quotation}

To further highlight the metaphysical nature of transduction, Simondon argues that transduction cannot be captured by the logical operations of either deduction or induction. Transduction is not deductive since it does not posit a given principle(s) or pattern(s) external to the process that can instruct the resolution of the present situation. Deduction can only highlight that which is already given by fully individuated knowledge. Transduction `discovers', or rather brings forth, elements and relations that did not exist before. Furthermore, transduction is not inductive in the sense that it does not extract or highlight the properties or patterns common to the unique and not yet compatible elements of the individuating process. These usually serve as the basis to inductive reasoning about the process, thereby eliminating what is unique to the elements. Instead, ``[T]ransduction is, on the contrary, a discovery of  dimensions of which the system puts into communication [...] each of  its terms, and in such a way that the complete reality of each of the terms of the domain can come to order itself  without loss, without reduction, in the newly discovered structures.'' \citep[p. 12] {simondon_position_2009}. 

\subsubsection*{Application to quantum phenomena}
In brief, quantum systems prior to measurement are not fully individuated. The measurement of a complementary property in a quantum entangled system is an \textit{individuating event} in respect to the property being measured\footnote{Generally, a measurement is not always an individuating event. It depends whether the actions involved in the measurement produce for the measured system a perturbation strong enough as to move it from its current stability towards another stability. But it can be said that in every individuation event, certain determinations must take place and therefore it can always be understood as measurement in the broad sense.} in the sense that it determines something that was not determined before, it brings forth a distinction, a differentiation. Measurement does not merely change the state of our knowledge about reality. It actually changes the state of both \textit{knowledge and reality}. As we have seen, these, according to Simondon, individuate together \citep{combes_gilbert_2013}. In this sense, measurement in quantum systems realizes the transduction mechanism.

Describing measurement as an individuating event elegantly fits the fact that complementary properties cannot be simultaneously measured and require distinct samples (see subsection \ref{subsec:bell_ineq}). Individuation takes place when some property which was not determined, gets determined. But clearly a property cannot be determined twice from the very same predetermined state. Individuating events are ultimately unique, hence can each be sampled only once. 

Interestingly, the concept superposition of states can be understood as a projection of individuated properties post measurement back to the non-individuated state of affairs prior to measurement. The wave functions being superimposed, are always in conjunction to an arbitrarily selected specific measurement (e.g. measuring spin or polarization in direction $ x $), they have no meaning independent of the measurement settings. This quantum wave modality illustrates best a system in the course of individuation. Even when an individuating event takes place, the system becomes individuated and yields a concrete and consistent outcome only in the context of that same event (measurement). But since every arbitrary measurement that follows potentially changes the reality of the system, the preindividual intrinsic to the system is never exhausted. 

A quantum system is therefore an exemplar of a metastable individuating system. All its individual products are always given only in relation to the latest individuation event. The probabilities associated with superimposed states should not be interpreted as if they reflect frequencies of already defined properties (like in the urn example). They rather indicate a statistical regularity of how the undifferentiated state might evolve and this depends of course on what is already known and what will be measured next.   

In summary, Simondon's theory presents a paradigm shift in the way we can relate to the quantum world: from a view based on individual entities, to a view based on ontogenetic processes that bring forth individual entities. The implications of this shift on understanding quantum phenomena are discussed next.
  	
\subsection{Individuation and realism} \label{subsec:individuation_realism} % Covers step 2-3
The most counter intuitive and intriguing behavior of entangled systems is the case where separability is challenged. Separability means that spatially separate systems posses separate real states (real is said here in the physical sense). \citet{howard_einstein_1985} strongly emphasizes the profound significance of separability for physics:
\begin{quotation}
	[I]t should be understood that the separability of two systems is not the same thing  as  the  absence  of  an  interaction between them, nor is the presence of an interaction the mark of their non-separability. The  separability  principle  operates on a more basic level as, in effect, a principle of individuation for physical systems, a principle whereby we determine whether in a given situation we have only one system or two. If two  systems are not separable, then there can be no interaction between them, because they are not really two systems  at all.
\end{quotation}
Quantum entangled systems definitely do not follow this principle. For example, a pair of particles\footnote{Generally, more than two particles can be entangled and form systems where individuation can take many paths and can become intractably complex.} having their spins entangled form a system which is not fully individuated and therefore inseparable. Distinct individual spins do not exist for each of the particles\footnote{We can call them particles because they may still posses other properties such as mass or charge that partially identifies them as distinct entities.} constituting the system; there is only an internally correlated joint spin state for the whole system. Of course inseparability is reflected in the mathematical formalism used to represent such states. It uses the principle of superposition borrowed from the fact that individual particles behave also as waves that can be superimposed and yields an expression that provably cannot be decomposed into separate expressions for each particle. 

Einstein's concerns regarding quantum theory were centered on the fact that in its very formalism it denies the principle of separability. For him, separability seemed to be the essence of realism: 
\begin{quotation}
	However, if one renounces the assumption that what is present in different parts of space has an independent, real  existence, then I do not at all see what physics is supposed to describe. For what is thought to be a ‘system’ is,  after all, just conventional, and I do not see how one is supposed to divide up the world objectively so that one can make statements about the parts. \citep[p. 191]{howard_einstein_1985} 
\end{quotation}
Einstein was worried that without separability, there will be no way to objectively distinguish between physical systems and this will inevitably leave us only with subjective, observer-dependent (and arbitrary) interpretation of what constitutes a physical system. It is also clear that Boole's conditions of possible experience are exactly those conditions under which the ``statements about the parts'' mentioned by Einstein can be safely made and tested. They are constructed in a manner that ensures the separability of observed systems. This is why we cannot expect entangled quantum systems to follow Boole's conditions because these require that all properties to be fully differentiated and separable (e.g., in the urn example, woodenness should never depend on redness etc.).

It is here that Simondon's metaphysical scheme becomes relevant to the problems discussed in this paper. What Simondon's scheme allows is to metaphysically accommodate individuating non-separable systems. This necessarily changes the whole view about reality (not only quantum reality): individuals occupy only a small and secondary part of reality. Entities in the course of individuation with yet undifferentiated properties are the rule rather than the exception. It is only an epistemological convention (and convenience) that we approximate such entities by representative individuals. Such approximation allows to represent phenomena in terms of discrete predicates and logical propositions. But reality is far from being fully captured by such representations, and apparently there are vastly more possible experiences than those allowed by Boole's conditions that apply only to individuals. 

We can contrast now \textit{hard realism} -- a description of phenomena in terms of fully formed individuals with \textit{soft realism} -- a much broader description that includes partially formed individuals with as yet undifferentiated properties\footnote{The term \textit{partial identity} can be synonymously applied.}. Applied to quantum phenomena, with soft realism we depart from the conventional realist position which is hard realism but we do not have to to go as far as the non-realist position that denies altogether the existence of measurement-independent properties either. To assert that measurement is instrumental to the individuation of certain systems is to affirm that reality is not something which is either \textit{a priori} given or does not exist at all but rather that reality is in a continuous process of individuation (ontogeny).  

We already know that representations based on hard realism are problematic. The violation of Boole's conditions is clear enough evidence for that. But does soft realism help us to achieve a more consistent representation? It seems that it does. Based on Simondon's metaphysical scheme, soft realism allows a novel kind of possible experience -- a partially individuated entity as exemplified by entangled systems. For such systems, the so called violations predicted by the theory and validated by experiment are not violations at all. Quantum theory predicts with unprecedented success the outcome of measurements performed on systems that are only partially individuated. The troublesome correlations discussed in section \ref{sec:EPR} positively indicate the inseparability of the entangled system, but now we have a metaphysical scheme that accommodates this fact. Physics remains intact and our understanding of the world gains a profound refinement and much needed consistency of representation. This view is supported by Howard as well:
\begin{quotation}
	[...] We should make the existence of quantum correlations a criterion of non-separability. After all, if  it  were not for the existence of these peculiar correlations which violate the Bell inequality, the separability  principle would not be threatened. In other words, what I suggest is that instead of taking the quantum correlations as a puzzle needing explanation, we should make these correlations themselves the explanation [...] \citep[p. 198]{howard_einstein_1985}
\end{quotation}
What Howard was seriously missing is the metaphysical backup provided by Simondon's theory. Without it, his suggestion seems to be merely an arbitrary choice of convenience. But it makes much sense in the light of the metaphysical scheme of individuation: spatial separation is not enough as the ultimate criterion of separability. We will discuss this further in section \ref{sec:space}.  

\subsection{Individuation and conditions of possible causal explanation} \label{subsec:causality} % covers step 3 

Individuation is an abstract process that does not provide the specific physical mechanism of the actual determinations and differentiations that take place in its course. For that matter it does not even provide a hint as to what kind of explanations one can expect. Quantum theory is a theory of statistical regularities. Conventional classical thinking seeks to explain statistical regularities as originating from actual distributions of properties in an hypothetical population of individuals with \textit{a priori} defined identities. But this approach is rooted in hard realism and does not offer a viable resolution of the paradox. From the perspective of soft realism, quantum theory describes systems undergoing individuation. These do not `hide' mysterious individual elements (hidden variables) on which causative explanations can be anchored. Does soft realism offers an alternative to conventional causative explanations? 

We have already seen in subsection \ref{subsec:realism} that abandoning realism might be a way out. Pitowsky makes it clear that the issue at stake is not so much giving up realism but the idea of giving up causality -- more precisely the principle that no event happens without a cause. Pitowsky's suggestion that the statistical regularities in the case of quantum entanglement have no causal explanation is not as speculative and dismissive of causation as it might seem at first sight. He writes: ``There is no 'deeper reality' which causes them [the statistical regularities] to occur; the phenomena themselves are their deepest explanation.'' \citep[p. 118]{pitowsky_george_1994}. If we carefully reexamine this quote in the light of the inseparability of entangled systems we find a deeper sense. Given two entangled particles $A$ and $B$, how can we describe the effect of measurement on $A$ on the measurement on $B$ if the particles are not separable? If $ A $ and $ B $ are one and not two distinct systems, in what sense can one produce effects on the other? Is it not the case that  both cause and effect are internal to a single non-decomposable whole and this is the best one can do in describing what is going on without dismissing causality altogether? 

Thinking in terms of individuals necessitates that in order for one entity to act upon another one and cause an effect, they need to be separate and external to each other. Even feedback systems that when observed from outside can be seen as if acting upon themselves, can always be represented as having internal structure that separates input subsystems from output subsystems. In this sense, a causative relation is always a relation of externality -- external to the related elements. Thinking in terms of individuation is entirely different. A system in the course of individuation is in a state where elements are not entirely differentiated yet not entirely homogeneous and indistinct either. In such systems a causative relation can be understood only as a relation of internality where both the acting and acted upon elements are not entirely distinct; their relation therefore is internal to them\footnote{The idea of the difference between relations of externality (that require the separation of elements) and relations of internality (that require continuity and interpenetration of elements) originated in the works of another eminent philosopher of beginning of the 20th century Henri Bergson \citep[227]{bergson_time_2001} (see also: \citep{deleuze_bergsonism_1991}). Bergson's work predates Simondon's and deeply inspired his philosophy of individuation. His work will be further discussed in section \ref{sec:space}}. And since they are not entirely distinct one cannot even discern the direction of action -- which of the elements is the acting and which is the affected\footnote{With relativistic considerations taken into account, the measurement on either particle can precede the measurement on the other depending on the frame of reference of the observer. In as far as causes precede effects the ambiguity of the situation is very real.}. In other words,  causes and their effects are confused. It seems that this state of affairs can receive only an approximate description using a language optimally fitted to mostly describe relations of externality. It can be said however that the causation relation itself is individuating and not entirely distinct (see above p. \pageref{relations} on the preindividual). 

The traditional concept of causality involved two requirements: spatio-temporal contiguity and regularity (similar causes are followed by similar effects) \citep{ben-menahem_struggling_1989}. The core of Hume's skepticism regarding causality was that the causing agent and the affected agent are ultimately distinct. There is always something that must come between them to mediate action and this necessity, Hume argued, cannot be logically established; it is only empirically established. In other words, effects cannot be logically derived from causes only inferred. In individuating systems we face a different and in some sense an opposite problem where the causing agent and the affected agent though spatially separate are not entirely distinct. Contiguity in this case attains a sense which is other than the traditional spatio-temporal relation; it is contiguity defined in terms of an additional property dimension. If this dimension represents for example the direction of spin, there is only a single (yet arbitrary) value representing the direction of both particles. The particles therefore are found contiguous on this dimension. In contrast to being a mere empirical fact obtained by observation; it is an ontological and logically established contiguity. Nowhere and in no case can one intervene to change the property of one particle without affecting a change in the other. If two things cannot be separated, they are necessarily contiguous in some very significant sense. It can be said, therefore, that the relation of entanglement is stronger than the traditional causal relation; the mutual effect is more profound. It is not mere regularity that we observe in the behaviors of entangled particles it is logical necessity arising from their non-separability. 

Furthermore, there is neither metaphysical nor logical reason to privilege one physical property (spatial separation) over another (non-separable spin states) in judging the distinctiveness of elements of an entangled system. Hence, `action over distance' fails to describe what is going on in entangled systems. Clearly the relation of entanglement involves both more than and less than what we conventionally conceive in the concept of action (causing something to happen). The following section will continue to further scrutinize the application of the notion of spatial separation to entangled systems. 

The position of soft realism towards causality again takes advantage of the concept of individuation to establish that physical interactions can be more subtle and complex as to neatly fit into or be excluded from the traditional category of causal relations. Individuating relations are understood as relations of internality rather than relations of externality. Measurement as an individuating event, externalizes (exposes to the external observer) a relation that was internal up to that point. Consequently, the states of both knowledge and reality have thus changed.       

To this point, the discussion focused on a single system of entangled particles. How does this analysis reflect on explaining the actual statistical regularities predicted by Quantum theory? Are we still stuck in a position that forces us to choose between a paradoxical explanation (non-local effects) and no explanation at all (non-realism)? Is there an alternative supported by the analysis above? There is no doubt that the statistical results of EPR type experiments are reflecting the behavior of a population of entangled systems. We can now see that the paradox arises because the population is of individuating entities and not of individuals. Since measurements are individuating events, and no single system can be individuated more than once, complementary properties must then be measured on distinct samples of the population. This wouldn't normally pose a problem if not for the fact that the population prior to measurement and the population post measurement are not the same populations. \textit{There is an metaphysical difference between the members of the two populations, they are not of the same kind}. Whereas the first is of individuating pairs, the second is of pairs of individuals. The paradox arises when we expect the first to behave as the second would as if they were of one and the same kind. Accepting this difference brings us back to Pitowski's words: ``[T]he phenomena themselves are their deepest explanation.'' If individuation is a fundamental state of reality there is no need to seek for a deeper explanatory element. It is entanglement itself that explains the correlations discovered in experiment. 

In summary, while conventional causative relation cannot be said to exist between the particles of a single entangled pair and needs to be replaced by the more refined understanding of their relation as suggested above, accepting entangled systems as individuating instead of individuals is enough to cause the observed statistical regularities. It is simply not an \textit{a priori} distribution but how individuation works. For any sample of the population, the fact of having belonged to a single undifferentiated entity leaves a trace in the behavior of each and every pair that measurement brings forth.      

\section{The individuation of space} \label{sec:space}
\subsection{Rethinking locality and its role in individuation} % Covers step 4
We have seen the criticality and problematics of separability to the understanding of entangled systems and to the notion of realism. The definition of separability as discussed in subsection \ref{subsec:individuation_realism} requires that spatially distinct system must have separate real physical states. As much as the definition seems simple and straight forward, a closer examination exposes an unexpected complication. If we understand spatial separation to be a purely physical property, there is no reason (as already mentioned) to privilege it over other physical properties in judging whether two physical systems are separate or not. Perhaps it is only a perceptual habit to see spatial separation as some kind of an primal criterion? Perhaps a pair of entangled particles is a single system spatially extended but spatial separation is only secondary in significance? More relevantly to our issue, we must consider systems where elements are both separable and not. If on the other hand space is more than just a pure physical property; that it somehow transcends the purely physical, than on account of such transcendence, its special privileged status might be justified. According to Kant, space indeed enjoys a special status:

\begin{quotation}
Space is not an empirical concept which has been derived from outer experiences. For in order that certain sensations be referred to something outside me (that is, to something in another region of space from that in which I find myself), and similarly in order that I may be able to represent them as outside and alongside one another, and accordingly as not only different but as in different places, the representation of space must already underlie them. [...] Therefore, the representation of space cannot be obtained through experience from the relations of outer appearance; this outer experience is itself possible at all only through that representation. [...] Space is a necessary a priori representation that underlies all outer intuitions. One can never forge a representation of the absence of space, though one can quite well think that no things are to be met within it. It must therefore be regarded as the condition of the possibility of appearances, and not as a determination dependent upon them, and it is an a priori representation that necessarily underlies outer appearances. \citep{kant_critique_1998}
\end{quotation}

In other words, the separability in space, which comes \textit{a priori} to any representation is a sufficient condition that systems separated by space alone are already \textit{physically separated} in any possible experience and therefore must also posses separable real states. \citet{janiak_kants_2012} makes an interesting distinction between a realist relationalism and realist absolutism in regards to space. Whilst the first is the position that space is the order of possible relations among objects, the latter is the position that space is an object-independent framework for object relations. From the perspective of the first position, it is conceivable that spatial separation might depend on other relations between objects (e.g. their quantum states). It might be the case that separation and conventional distance are not one and the same and not any conventional distance automatically reflects a separation as this may depend on other non-spatial relations between the systems under consideration. The second position that claims an object-independent status to space, seems however to be the one consistent with Einstein's views. From the standpoint of special relativity, signals can move through space-time only in a limited speed. Spatial separation means therefore a limit on the communication between two physical systems and this limit was in Einstein's eyes a fundamental one because this very communication is \textit{a priori} intrinsic to \textit{any relation} and any physical interaction between two physical entities. 

Yet it is clear that entangled systems that are spatially separated do not communicate in a manner that violates special relativity in any respect. They do however \textit{relate} in a special manner as if no spatial separation exists between them. I argue here that Einstein's concern regarding quantum phenomena  arose from his realist absolutism position which is a metaphysical one\footnote{It might sound strange that Einstein who conceived relativity theory held a realist absolutist position. But the sense of absolutism here is the claim that spatio-temporal relations between objects are the basis to any other relation and antecedent to any other relation. `Realist absolutist' is just another name to a position that privileges spatio-temporal distinctions to any other distinction.}. But there is no compelling point to hold to this position because it is not the only one that is consistent with empirical data. The alternative realist relationalism which is intrinsic to soft realism is consistent with existing theory and all empirical data and in the case of the EPR paradox invites to reexamine the deeper meaning of locality in entangled systems and whether the principle of locality is indeed violated. I will argue that based on the metaphysical scheme of individuation, in the case of entangled objects, space as the order of possible relations among objects is itself subject to individuation inasmuch as the relations it orders themselves individuate. This will lead us to consider an additional metaphysical adjustment having to do with the concept of locality. The argument is based on the metaphysics of space  developed by Bergson which is briefly presented next. 

\subsection{Bergson's metaphysics of space} \label{subsec:bergson} % Covers step 4
Bergson's metaphysics of time and space is very rich and complex. It is not within the scope of this paper to provide the wider context of Bergson's writings\footnote{Especially ``Time and Free Will'' and ``Duration and Simultaneity'' -- Bergson's investigation of relativity theory \citep{bergson_time_2001,bergson_duration_1965}.} which are necessary for the deeper grasp of his metaphysical method. Here we try to extract in brief only the few points which are relevant to the topic at hand. At the basis of Bergson's thought about the metaphysical nature of space is a combination of the following three philosophical observations.

\subsubsection*{Distinction between space and extensity}
In \citep{bergson_time_2001} we find the following:
\begin{quotation}
	We must thus distinguish between the perception of extensity and the conception of space: they are no doubt implied in one another, but, the higher we rise in the scale of intelligent beings, the more clearly do we meet with the independent idea of a homogeneous space:
\end{quotation}
The distinction is a subtle one: while extensity is an actual objective manifestation of physical objects, space, Bergson argues, is conceptual and involves the mind of an observer. The exact nature of the concept and its function will become clear in the following.  
\subsubsection*{Homogeneity of space}
Furthermore, Bergson contrasts the qualitative heterogeneity of our conscious experience with the homogeneity of space:
\begin{quotation}
	What we must say is that we have to do with two different kinds of reality, the one heterogeneous, that of sensible qualities, the other homogeneous, namely space. This latter, clearly conceived by the human intellect, enables us to use clean-cut distinctions, to count, to abstract, and perhaps also to speak. \citep[p. 97]{bergson_time_2001}
\end{quotation}
According to Bergson, the reality of experience is a continuum of heterogeneous qualitative change that does not admit any intrinsic distinction or separation. Only by projecting this continuum onto space, one can start making distinctions and separations: 
\begin{quotation}
	[S]pace is what enables us to distinguish a number of identical and simultaneous sensations from one another; it is thus a principle of differentiation other than that of qualitative differentiation, and consequently it is a reality with no quality. \citep[p. 95]{bergson_time_2001}
\end{quotation}
The homogeneity of space is exactly this: being devoid of quality. As such, it is always external to anything with quality. Therefore, space is the kind of reality that enables relations of externality. Without applying the concept of space one can only conceive of relations of internality where no separation can be made\footnote{This is quite easy to see: if something having a quality $ A $ changes into having quality $ B $ what happens at the limit between $ A $ and $ B $? The limit must either consists of both $ A $ and $ B $ or neither, for any of the other options (i.e. either $ A $ or $ B $) is not consistent with it being a limit. If the limit consists of both $ A $ and $ B $ it is impossible to fully separate $ A $ from $ B $ because at least at their limit they are inseparable. The option that the limit consists of neither $ A $ nor $ B $ is indeed the only one left. Now suppose the limit consist of having another quality $ C $ different from both $ A $ and $ B $, then we must now ask recursively the same questions about the limit where $ A $ changes to $ C $ etc. We are left therefore with the option that $ C $ is the absence of any quality. Only in such case we can claim that $ A $ and $ B $ are indeed mutually external to each other and entirely separate. $ C $ is Bergson's conception of space. It is in fact the Aristotelian middle being excluded.}. 
\subsubsection*{Space is infinitely divisible} 
The third and most important observation is brought in the following quote from Matter and Memory \citep[p. 206]{bergson_matter_1991}: ``Abstract space is, indeed, at bottom, nothing but the mental diagram of infinite divisibility.'' Bergson further explains: 
\begin{quotation}
	Such is the primary and the most apparent operation of the perceiving mind: it marks out divisions in the continuity of the extended,  simply  following the suggestions of our requirement and the needs of practical life. But, in order to divide the real in this manner, we must first persuade ourselves that the real is divisible at will. Consequently we must throw beneath the continuity of sensible qualities, that is to say, beneath concrete extensity, a network, of which the meshes may be altered to any shape whatsoever and become as small as we please: this substratum which is merely conceived, this wholly ideal diagram of arbitrary and infinite divisibility, is homogeneous space. \citep[pp. 209-210]{bergson_matter_1991}
\end{quotation}		

\subsubsection*{Space and individuation}
 Bergson's thought brings forth interesting points relevant to our investigation. Understanding space as a mental diagram of divisibility and distinguishing it from extensity means that physical extensity does not automatically imply divisibility. In other words, physical objects and systems may be extended without being divisible. Moreover, divisibility is not fundamental; continuity and non-separation are the fundamental conditions of the real according to Bergson. Divisibility which is necessary for separability is not intrinsic to the real\footnote{Remakably, the concept of distance as reflecting spatial separation is applicable therefore only on account of space being homogeneous, devoid of quality and divisible. Without these, we can speak distance only as some conventional measure of extensity.}; it requires an extra ``ideal diagram'' to be casted beneath the real. This is to say that spatial separation based only on extensity \textit{is not} metaphysically privileged (or \textit{a priori} warranted) over other physical qualities. In fact, the very idea of pure spatial separability that appears to be deeply intuitive is put into question.
 
 One may go as far as concluding that Bergson's concept of space and divisibility is fundamentally subjective and requires the intervention of the mind of an observer. But Bergson's idea is more subtle: divisibility and separation may still be observer independent thus sustaining their realist status. However, one cannot distinguish between two objects \textit{only on account of the absence of a quality}. Space can be thrown beneath a continuity of sensible qualities, but if these are missing there is nothing to divide or separate. In other words, physical entities, whether observed or not, are not and cannot be separated only on account of purely geometrical relations. 
 
 From here it is clear how this conceptualization of space is consistent with the realist relationalism position mentioned in the previous subsection: space as the order of \textit{possible relations} among objects. In as far as physical entities can be separated at all, and since they cannot be separated only on account of an absence of quality (i.e. only spatial separation), it follows that the condition of their separation is that their possible relations must be based on concrete qualities/properties. There must be something rather than nothing which space might divide. But if that `something', that physical entity, is spatially extended but nevertheless intrinsically indivisible, space cannot possibly make it divisible though our habitual intuitions may tell us otherwise. Only in cases where physical entities are divisible and separable on account of other qualities or properties, the idea of space applies to represent their distinction. At the beginning of subsection \ref{subsec:individuation_realism} we defined and discussed the concept of separability (i.e. spatial separation implies separation in state) and its importance. It is clear now that the concept is based on presuming the primacy of spatial separation over all other quality based distinctions. This primacy, we find, is merely a feature of a particular metaphysical scheme. I have presented here an alternative metaphysics of space without this particular feature. I also argue that there is no reasonable basis for such primacy. The bottom line of this whole discussion is that in our proposed metaphysical scheme, spatial separation \textit{does not and cannot imply} separation in state on its own; on the contrary, separation in state\textit{ is a necessary condition} to spatial separation.       
 
 From the perspective of the metaphysics of individuation, there is no \textit{a priori} condition of entities being divisible or not. In other words, divisibility (i.e. separability) itself can individuate; which means that it is possible that a certain physical entity is spatially extended but not divisible will individuate and bring forth two or more spatially separate entities that did not exist before\footnote{Also the other direction is possible: where two or more spatially separate entities merge into a single undifferentiated spatially extended entity.}. This very possibility of the individuation of separability as a consequence of the individuation of other physical states and relations is the additional adjustment we need to accommodate following the metaphysical scheme of individuation.  
 
\subsection{Locality in entangled systems}  % Covers step 5
Let us now return to the case of a system constituted by two entangled physical entities in an EPR kind of setup. Following our new metaphysical scheme, though the system is spatially extended, we cannot take for granted anymore that the entangled entities involved are spatially separated. Considering \textit{only} the entangled property, there is no way we can assign an independent state to any of the entities constituting the system and this implies that space as a principle of differentiation is not applicable. In other words, \textit{there is no meaningful way to speak about locality or distance within the entangled system}. This may seem quite incredible and counter intuitive but this is only because our profound habit of perceiving and thinking in terms of sharply defined individuals and also that anything spatially extended is also spatially divisible. 

We already argued in subsection \ref{subsec:causality} that the spooky action at a distance that has become emblematic of entangled systems is not an action in the conventional sense of the word. We now complement that argument: it might well be a spooky action but \textit{there is no distance that reflects distinct localities}! It is by now a well established fact that entangled systems behave the same no matter how far they are spatially extended. What we conventionally measure as a distance between the entangled entities reflects extension but the separation it seem to reflect is only a feature of our conditioned imagination. The physical entities involved do not behave in any manner as to indicate that there is any spatial separation derived from the measured distance between them. 

Reality, of course, is not that simple. Each of the entities constituting an entangled system may (and often does) possess in addition to the entangled property other properties such as mass and charge that are independent. In such cases which are the majority, the spatial separation between the entities is not fully individuated. The entities are \textit{both spatially separable and inseparable} depending on which of their properties is under consideration. Such state of affairs which is contradictory according to the metaphysical scheme of individuals, is entirely consistent within the metaphysical scheme of individuation.              

In summary, locality in entangled systems is not an \textit{a priori} given but individuating. Prior to measurement, it is not yet differentiated. The system is spatially extended by not spatially divisible. A measurement of the entangled property is an individuating event; it brings about a differentiation in state and consequently spatial separation between the once entangled entities.

Inasmuch as individuation allows us to think in terms of soft realism, it allows us to think in terms of soft locality too. Hard locality is based on the assumption that spatial separability is a given and therefore the location of a physical entity in relation to other such entities can always be singled out. Soft locality does not assume that; instead, it accepts that spatial separability is not a given and is not applicable in the absence of other separating physical properties. Contrary to that, it is conditioned on the presence of entities with independent properties or states. If physical entities do not possess such properties they cannot be said to be spatially separated even though they together may constitute a spatially extended system. Soft locality does not give up locality but distinguishes between cases where space as a differentiating principle is applicable and cases where it is not. But most remarkably, soft realism accepts locality as an individuating feature of physical systems.  

% only when a system individuates divisibility is actualized! Space is not automatically divisible in actuality only virtually.
% spatial separation is never alone a separation it is always the distribution of another property in space (extension) so, space itself indeed does not separate anything, it is only the distribution of other properties in space that separate...
% Divisibility of space is ALWAYS complementary to some other physical property and its distribution. But also distribution is a matter of divisibility.
% if an entangled pair is a single location whatever the distance can we relate to it as an individual? 

\section{Conclusion} %covers step 6
The problematics presented by the EPR paradox have nothing to do with the facts of physics and the predictions of quantum theory. We have clarified that the problem is first and foremost epistemological. We expect that our knowledge of reality to be coherent and consistent across all phenomena at all scales but the observed violations of Boole's conditions of possible experience in the behavior of quantum entangled systems clearly put in question such coherency. 

Along almost a century of discourse about how to resolve the paradoxical findings involved in quantum phenomena, three different ideas played a major role namely: 
\begin{inparaenum}[\itshape a\upshape )]
	\item the incompleteness of quantum theory,
	\item accepting non-locality, and
	\item accepting non-realism. 
\end{inparaenum}   
Though we can never assure the completeness of the theory, the latest empirical findings prove that even if quantum theory is incomplete, either non-realism or non-locality are still necessary to resolve the apparent paradox. Clearly, neither of these resolutions coexists comfortably with how we understand the rest of reality. The more we try to cohere them the more disturbing they become. 

In this paper I propose a different approach to the resolution of the paradox. It is argued that the paradox is rooted in the metaphysical scheme that is supposed to provide a foundation to our knowledge of reality (including also our intuitions). This metaphysical scheme is based on the idea that reality is given in terms of individuals. It is from this idea that Boole's conditions of possible experience are derived. I argue that the violations of Boole's conditions do not indicate neither an incompleteness of quantum theory, nor lack of understanding its meaning. What they do indicate is failure to accommodate certain phenomena within the metaphysical scheme that we use. In other words, \textit{there are} actual experiences that do not comply with Boole's conditions which means that they cannot be described within a metaphysical scheme based only on individuals. The solution to the paradox would therefore be achieved by modifying and expanding the metaphysical scheme that we use as a basis of our knowledge of reality into one that can accommodate the actual experiences involved in quantum phenomena. 

Based on the works of Simondon and Bergson, I have proposed here an alternative metaphysical scheme which is based on the idea that reality is given in terms of processes of continuous individuation and where individuals are only impermanent products of such processes. I have shown that within such a scheme our notion of realism changes from hard realism to soft realism and also, as space itself individuates, our notion of locality changes from hard locality to soft locality. I have further shown that individuation as a metaphysical concept and the consequent adjustment thus made to both realism and locality, allows us a coherent description of quantum entanglement within a wider epistemological framework and provides an elegant resolution of the paradox.

As a concluding note I would like to briefly reflect on the method underlying this paper. Though the discourse in this paper focused on the EPR paradox, most of the arguments that were brought here and the application of the metaphysical scheme of individuation are relevant and generally applicable to a very wide spectrum of phenomena. It is my belief that metaphysics has an important if not critical role in the progress of scientific knowledge and our general understanding of reality. Metaphysics systematizes a set of fundamental assumptions about reality. Surely, it does not precede reality but it does precede the manner by which we perceive and conceptualize our perceptions into representations that constitute a coherent and reliable body of knowledge. As such, and because it is so fundamental it tends to hide from the scrutinizing eye and its precepts are often taken for an unassailable truths never to be questioned. There is only one remedy to this situation: metaphysics must not be isolated from science and science must not be isolated from metaphysics, one must keep the other in progressive check as both are undergoing individuation. Together they form a vital exchange, a dance that brings forth knowledge and the elegance of coherence.     
 
%Did we `save' epistemology after all?
% see conclusion of pitowsky
% we do not save physics we save ourselves by refining the scheme of how things are. This testifies the importance of ontology.
\pagebreak
\printbibliography

\end{document}